\documentstyle[11pt]{article}
\newcommand{\beq}{\begin{equation}}
\newcommand{\enq}{\end{equation}}
\newcommand{\ber}{\begin{eqnarray}}
\newcommand{\enr}{\end{eqnarray}}

\newcommand{\eq}{equation }

\newcommand{\er}[1]{\eq (\ref{#1}) }

\title{Cortical Dynamics and Awareness State:
An Interpretation of Observed Interstimulus Interval
 Dependence in Apparent Motion.}

\author{R. Englman \\Soreq NRC, Yavne 81800 \\
e-mail: englman@vms.huji.ac.il \\ and \\
A. Yahalom \\ Racah Institute of Physics, The Hebrew
University \\ 91904 Jerusalem, Israel\\
e-mail: asher@yosh.ac.il}

\begin{document}

\maketitle

\begin{abstract}

 In a recent paper on Cortical Dynamics, Francis and Grossberg raise
 the question how visual forms and motion information are integrated
 to generate a coherent percept of moving forms? In their investigation
 of illusory contours (which are, like Kanizsa squares, mental constructs
 rather than stimuli on the retina) they quantify the subjective impression
 of apparent motion between illusory contours that are formed by two
 subsequent stimuli with delay times of about $0.2$ second (called the
 interstimulus interval ISI). The impression of apparent motion is due to
 a back referral of a later experience to an earlier time in the conscious
 representation.

 A model is developed which describes the state of awareness in the observer
 in terms of a time dependent Schroedinger equation to which a second order
 time derivative is added. This addition requires as boundary conditions the
 values of the solution both at the beginning and after the process.
 Satisfactory quantitative agreement is found between the results of
 the model and the experimental results.

 We recall that in the von Neumann interpretation of the collapse of the
 quantum mechanical wave-function, the collapse was associated with an
 observer's awareness. Some questions of causality and determinism that
 arise from later-time boundary conditions are touched upon.

\end{abstract}

\section{Experimental Background}

 We consider  phenomena in cognitive psychology variously  known as
 "apparent movement" \cite{fr}, temporal "binding"
 (\cite{tr}, \cite{fa}),"filling in" \cite{ar} or
 "coherence" \cite{gr}.

 It was noted several years ago that when a bright spot on a screen was
 followed (after an interval of up to half a second) by another spot some
 distance away, the viewer perceived the spots as though performing a
 continuous motion \cite{ko}. The time delay involved is much longer
 than the characteristic delay or inertial time (tens of milliseconds)
 in cortical  neural cells and it is regarded as due to the time needed
 for conscious processing. Moreover, having perceived the second spot,
 the viewer does not "date" the motion to the instant of her seeing this,
 but to the instant of observation of the first spot. In the words of an
 authoritative source \cite{ve} [with added context-setting by us]:
 "... there is a subjective referral of timing \cite{li}, i.e.
 the brain compensates for [the] preconscious processing [what happens
 before $t=0$] by "marking" the time of arrival of stimuli at the cortical
 surfaces [at $t=0$] with an evoked potential and then referring experienced
 time of occurrence [$t=T$] back to time of arrival of stimuli
[$t=0$].
 Therefore experienced [!] time of arrival is actual time of
arrival [$t=0$],
 rather than the time when a conscious representation has had time to
 develop."

It is reasonable to assume that a quantification of awareness
variables \\
 (the "qualia") is feasible and access to them exists through psychological
 and neurobiological inquiries (\cite{de}, Section 4.3).
 Arguments have also been made for the existence of awareness units,
 analogous to atoms in matter. \cite{po} Estimates of the speeds of
 mental processes, e.g. decision times  ($0.06-0.17$ s) have been
 obtained in
 the classic studies of Donders \cite{do}. These are distinct from the times
 needed for "neural conditions to develop that are adequate to support
 conscious experience" \cite{ve} and presumably relate to processes in
 upper layers of the cortex.

\section{Wave-function formalism}

We describe cognitive processes in terms of brain ($r$) and {\bf
awareness}
 ($A$) variables and assume that their combined state
$\Psi(r,A;t)$ ($t$ is time) is subject to the formalism of wave
mechanics.
 This hinges on a Hamiltonian $H(r,A; t)$ which prescribes the evolution of
 the state of cognition. Since, however, Schroedinger equation is completely
 causal (in the sense that its solution evolves in time in a unique manner),
 whereas mental processes are (by our understanding) non-deterministic, we
 write the following equation for  $\Psi(r,A;t)$.
\beq
-W \frac{\partial^2 \Psi}{\partial t^2}+ ih \frac{\partial
\Psi}{\partial t} = H \Psi. \label{ce} \enq
Since this is a {\bf second} order differential  equation in time,
its solutions
 are fixed by boundary conditions imposed on $\Psi$ both
 at the beginning and at the {\bf end} of the cognitive process.
 This type of
 boundary conditions is appropriate for the description of  human choices
 and decisions (which are {\bf felt} to be non-predestinated)
 and for the collapse
 of wave packets in the course of a quantum mechanical measurement
 (whose outcomes are also not predetermined). [We note that the original
 form of the non-relativistic limit of the Dirac equation also has a
 second order time-derivative, rather than only the first order derivative,
 as in the Schroedinger equation.\cite{pa}] In simple terms this means
 that in order to describe certain mental processes one has to know
(in addition to the "mechanism", expressed by the Hamiltonian) the final as
 well as the initial state. The probability (or frequency) of a final state
 actually occurring is given by Born's propensity rules.[Exactly as in
 quantum mechanics ( Probability $\propto $ weight of final
 state in the initial state).]

\section{Awareness state}

\label{aw}

    For constant H one can solve \er{ce} in the domain
 (called "Domain II") between $t-0$ and $T$, with boundary conditions
 imposed at the edges of the domain. For $\frac{HW}{\hbar^2} \gg 1$ [which is
 appropriate for brain sizes having macroscopic (not atomic)
 dimensions] one obtains solutions of the type:
\beq
\Psi= a \exp{t \sqrt{\frac{H}{W}}} +
 b \exp{-t \sqrt{\frac{H}{W}}}.
\enq
The norm of this state function is time dependent and has a minimum at
 the midpoint of the domain (at $t=T/2$) of about
\beq
|\Psi|^2_{min} = \exp{-T \sqrt{\frac{H}{W}}}.
\enq
Because of the conservation of all matter in the brain, the natural
 interpretation of this result is that awareness is {\bf not} conserved but
dips inside domain $II$. To test this in a specific, but still hypothetic
 way we turn to observation of apparent motion between illusory contours
 \cite{vo}. This phenomenon was recently quantitatively
investigated by Francis and Grossberg \cite{fr}, who obtained
the percentage of positive responses from two subjects (GM and
PG) as function of stimulus duration D (in the range
$0.04-0.64$ s) and of the interstimulus interval (ISI) in the
same range (Their figure 1(a)).
We identify the perception of motion with the arousal of
awareness and the ISI with our parameter $T$. We further
suppose that there is a threshold of awareness intensity given
by some value of the state function squared $l_0= |\Psi_0|^2$
below which the apparent motion will not be perceived
(i.e. the subject will say he has not seen any motion). For
simplicity we shall make a further assumption (similar to the
ergodic hypothesis); namely, that the fraction of times that
the subject will give a positive answer will be equal to that
fraction of time length within domain II during which the state
function intensity exceeds the threshold; or to the fraction of
time that
\beq
|\Psi(t)|^2>l_0.
\label{l0}
\enq
Approximating $|\Psi(t)|^2$ in the region $0<t<\frac{T}{2}$
by a simple exponential:
\beq
\exp{-2 v_+ t}
\enq
and in the region $\frac{T}{2}<t<T$ by its mirror image (so
that $|\Psi|^2$ is unity at $t=0$ and $T$), we obtain for the
fraction of time $f(T)$ that \er{l0}  is satisfied:
\ber
f(T) &=& \frac{\log{\frac{1}{l_0}}}{v_+ T}
\qquad {\rm if} \quad
\frac{\log{\frac{1}{l_0}}}{v_+ T} \le 1.
\nonumber \\
f(T) &=& 1 \qquad {\rm if} \quad
\frac{\log{\frac{1}{l_0}}}{v_+ T}>1.
\label{fT}
\enr
We wish to test this $f(T)$ vs. $T$ against the probability vs.
ISI values of Francis and Grossberg \cite{fr}. Neither $l_0$ nor
$v_+$ are known and indeed may vary with subject and $D$
(stimulus duration), but we see from \er{fT} that for a given
subject and $D$ $f(T)T$ is independent of $T$ or of $f(T)$.

\begin{table}

\begin{tabular}{|l|l|l|l|l|l|l|l|} \hline\hline
 Subject & D  & Value & 0.85 & 0.7 & 0.55 & Average & s.d. \\ \hline\hline
GM  & 160 & $ISI$ & 196 & 251 & 310 &  &   \\
    &     & $ISI \times Prob.$ & 167 & 176 & 171 & 171 & 4.5  \\ \hline
GM  & 320 & $ISI$ & 209 & 282 & 361 &  &   \\
    &     & $ISI \times Prob.$ & 178 & 197 & 199 & 191 & 11.6  \\ \hline
GM  & 640 & $ISI$ & 187 & 238.5 & 293 &  &   \\
    &     & $ISI \times Prob.$ & 159 & 167 & 161 & 162 & 4.2 \\ \hline
PG  & 160 & $ISI$ & 377 & 490 & 624 &  &   \\
    &     & $ISI \times Prob.$ & 320 & 343 & 343 & 335 & 13.3  \\ \hline
PG  & 320 & $ISI$ & 343 & 422 & 505 &  &   \\
    &     & $ISI \times Prob.$ & 291 & 295 & 278 & 288 & 8.9  \\ \hline
PG  & 640 & $ISI$ & 175 & 258 & 353 &  &   \\
    &     & $ISI \times Prob.$ & 149 & 181 & 194 & 175 & 23.2  \\ \hline
\end{tabular}
\caption{Values of Interstimulus Intervals (milliseconds) at
which subjects GM and PG perceived motion with shown
probabilities (relative frequencies) at given values of
stimulus duration ($D$, in ms). From Francis and
Grossberg \protect\cite{fr}.
Adjacent rows show the computed products
 $ISI \times Probabilities$  and their averages for each row and
the standard deviations (s.d.). The theory states that
$ISI \times Prob. $ is a constant in each row. }
\label{isi}
\end{table}

The following table (Table \ref{isi})exhibits the values for
\beq
ISI \times Probability
\label{Pisi}
\enq
at three equi-spaced values of $\log{D}$ and
for both subjects, taken from the contour plot of Francis and
Grossberg \cite{fr}. The constancy of the product in \ref{Pisi}
is indeed apparent. To better appreciate this, we show the
averages and the standard deviation (s.d.) of the product of
each case. The s.d. are much smaller ($11$ is the mean s.d.)
than the deviation between the averages (the maximum deviation
of averages in Table \ref{isi} is $170$ and the root mean
square of the differences between the $6$ averages is $106$).

\section{Conclusion}

    This work has started out with a simple idea that intended to fill
 in a gap in the conventional formulation of Quantum Mechanics, namely to
 describe the reduction of the wave-packet  as a continuous time-varying
 process. The second order differential \er{ce}, together with its
 boundary conditions, does this. Questions arising from it (e.g.,
 nonunitarity of development, causality) are not harder than those for
 previous works,e.g. the two-time formalism of  Aharonov and co-authors.
 An interpretation of the awareness state function has been discussed in
 section \ref{aw}.

In the awareness interpretation, the collapse is an expression of
 the mind-brain interaction and the collapse-equation proposed in this work
 is a formalization of this interaction. When we extend the reign of the
 equation to cognitive processes in general, we find novel implications in
 the following  fields:
 The psychological phenomenon of back referral (see section \ref{aw}), the
 quantification of conscious activity in small species , the process of
 mental decisions and the issue of free will .

 As the a next step, one needs to devise some critical
 experiments whereby the theory can be tested.

\end{document}